\documentstyle[preprint,aps,prc]{revtex}
 
\begin{document}
\draft 
\preprint{KSUCNR-101-98}
\title{Medium effect on photon production in 
       ultrarelativistic nuclear collisions}  
\author{Chung-sik Song\thanks{Electronic mail (Internet): 
csong@cnrred.kent.edu} and George Fai\thanks{Electronic mail (Internet): 
fai@cnrred.kent.edu}}
\address{Center for Nuclear Research, Department of Physics\\
         Kent State University, Kent, OH 44242}
\date{\today}
\maketitle
 
\begin{abstract}
 
The effect of in-medium vector and axial-vector meson masses on photon
production is studied. 
We assume that the effective mass of a vector
meson in hot nuclear matter decreases according to a
universal scaling law, while that of an axial-vector meson is given 
by Weinberg's mass formula. We find that
the thermal production rate of photons increases 
with reduced masses, and is enhanced by an order of magnitude at $T=160$ MeV 
with $m_\rho=300$~MeV. Assuming a hydrodynamic evolution, 
we estimate the effect of the reduced masses on 
photon production in nucleus-nucleus collisions.
The result is compared to experimental data from the WA80/WA98 
collaboration.                             
\end{abstract}
 
\pacs {PACS numbers : 25.75.+r, 24.85+p, 12.38.Mh}

\section{Introduction}

For hadronic matter at sufficiently high temperature and/or density, 
we expect a phase transition into a locally thermalized, deconfined plasma of 
quarks and gluons, the so called quark-gluon plasma (QGP)\cite{qgp}. Chiral 
symmetry, spontaneously broken in the ground state of quantum chromodynamics 
(QCD), is expected to be restored under such extreme conditions as shown by 
numerical simulations on the lattice\cite{lattice}. These properties of QCD 
are of great interest in order to understand the behavior of strongly 
interacting particles.  High energy nucleus-nucleus collisions offer a unique 
opportunity to explore the properties of QCD, in particular its phase 
structure, at high temperatures and densities\cite{rhic}.

Photon and lepton pair production have been suggested more than 20 years ago
as promising probes to 
study the properties of hot dense matter in high energy nucleus-nucleus 
collisions \cite{em_probe}. The strong temperature dependence of the production
rate of these signals makes them potential tools that can hopefully 
discriminate 
the various states of hadronic matter with 
different temperatures. In addition, these signals carry information 
on the hot 
dense matter without further distortion, since these electromagnetic probes 
interact very weakly with surrounding particles.
However, most model calculations show that the hadronic contributions 
dominate the signals from quark-gluon plasma \cite{dilepton,photon}. 
This implies that the bulk of the photon or dilepton spectra
observed in experiments will provide information on the 
properties of the hadronic phase under extreme conditions but 
below (or at most at) the phase transition temperature and/or density.

Recent interest in dileptons with low invariant mass has been sparked by 
the experimental data from the CERES collaboration \cite{ceres}, which show a 
considerable enhancement at invariant masses of about 500 MeV for SPS-energy 
S+Au collisions as compared to proton-proton and
proton-nucleus collisions. A similar enhancement has also been seen by the 
HELIOS collaboration at more forward rapidity \cite{helios}. It is noteworthy
that there is also an enhancement at energies above the $\rho$-peak
in these data\cite{rhic}. Suggestions have 
been made that the excess dileptons seen in these experiments are from 
pion-pion annihilation, $\pi^+\pi^-\to e^+e^-$. However, model calculations 
which have taken this channel into account can at best reach the lower end of 
the sum of statistical and systematic errors of the CERES-data in the low 
invariant mass region \cite{VKoch}. 
For the HELIOS results, which are unfortunately given without a  
systematic error, there is still a disagreement by up to 50\% between the
data and the calculations 
around  an invariant mass of 500 MeV \cite{VKoch}. It is, therefore, 
interesting to see to what extent medium corrections modify dilepton 
production. It has been suggested that the enhancement may be due
to the dropping of the vector meson masses at finite temperature and 
density \cite{gqli}, as proposed in Ref. \cite{br}, and/or to the broadening 
of vector meson resonances by the many-body effects in dense matter 
\cite{rapp}.
However, more 
conservative effects, such as a modified in-medium pion dispersion 
relation \cite{song_pion} or secondary mesonic scattering \cite{secondary}, 
may also give dilepton enhancement. 

In this paper we consider the medium effects on photon production at SPS 
heavy-ion collisions. In particular, we concentrate on the effects of dropping
vector and axial-vector meson masses on photon production. The result 
is compared to recent measurements of the WA80 collaboration which provide
an upper limit for photon production in hot and dense matter 
\cite{ex_wa80}, and to the preliminary results of the WA98 collaboration 
\cite{rhic}.
It has been shown that these experimental data can be well 
described by the hadronic contributions without any medium effect 
\cite{th_wa80}. Thus, it might be very interesting and valuable to see 
whether the medium effects which play an important role in dilepton 
production are also consistent with the photon spectra. 
Recently a calculation that incorporates self-consistently the 
change of hadron masses 
in dense matter has been carried out in 
a relativistic hadronic transport model \cite{gqli_photon}.
Non-thermal direct photon production has also been discussed recently in the 
UrQMD  framework \cite{urqmd}.

In the following section, we study the effect of the in-medium vector and 
axial-vector meson masses on the thermal production rates of photons. 
We assume that the vector meson mass is reduced according to a
universal scaling law and the axial-vector meson masses are given by
Weinberg's mass formula \cite{weinberg}. 
We include these medium effects in a dynamical 
model calculation which simulates the evolution of hot dense matter produced 
in nucleus-nucleus collisions in Section 3. We consider a one-dimensional 
hydrodynamic expansion for the hot dense matter with Bjorken's initial 
conditions \cite{bjorken}. We study
cases with and without a first-order phase transition.
The medium effect gives different initial temperatures, 
phase-transition temperatures, and mixed-phase lifetimes for the 
two different scenarios. In the Appendix,
we summarize thermal production rates for photons in
hot hadronic matter and present numerical parameterizations for 
the production rates for the use in dynamical calculations. 

\section{Photon production from hot hadronic matter}
    
Photon production from hot hadronic matter has been calculated by Kapusta
et al. \cite{photon} with a simple effective model for hadrons. This study
has been motivated by the need to decide whether the spectra of single photons
produced in high energy nucleus-nucleus collisions can tell us anything 
about the 
formation of quark-gluon plasma in the early stage of the collision. However, 
the authors found that ``the hadron gas shines just as bright as 
the quark-gluon plasma''. 
Moreover, it has later been pointed out that photon production from 
reactions including axial-vector mesons dominates that from the processes 
considered in Ref. \cite{photon} in hot hadronic matter \cite{lxiong}. 
The same conclusion has been obtained from an extended calculation 
based on a chirally symmetric effective Lagrangian with 
vector and axial-vector mesons \cite{song_photon}.
This implies that the photon 
spectra, especially for photons with $E_\gamma<3$ GeV, are dominated 
by the hadronic sources and provide information only on the hadronic phase in 
collisions. We summarize the results for various reactions and 
numerical parameterizations of them in the Appendix. 
The same conclusion holds true even when we include the space-time 
evolution of the hot and dense system produced in heavy-ion collisions 
\cite{th_wa80,expand}. 
However, the medium effects on the properties of mesons have not been addressed
in these calculations. 
 
In hadronic matter at high temperature and/or density we expect that 
properties of hadrons, such as effective masses and decay widths, are 
modified due to the phase transition/crossover into the deconfined phase of 
hadronic matter. 
Even below the phase transition temperature, chiral symmetry is 
expected to be partially restored, i.e. the magnitude of the order parameter
$<\bar q q>$ is reduced from its vacuum value. 
It has been suggested that the effective masses of vector mesons $(m_\rho)$
would be reduced in hot dense matter according to a universal 
scaling law as \cite{br} 
\begin{equation}
\frac{m^*_\rho}{m_\rho} \approx
\frac{m^*_N}{m_N} \approx
\frac{f^*_\pi}{f_\pi}    \,\,\, ,
\end{equation}
where $m_N$ is the nucleon mass, $f_\pi$ is the pion decay constant and 
$*$ indicates the in-medium value of the corresponding physical quantity.
These changes have significant impact on the hadronic observables as well as 
on electromagnetic probes \cite{br_ph}. 
Recently it has been shown that the enhanced low mass 
dileptons observed by the CERES experiment can be well described 
using a `dropping vector-meson mass' in medium \cite{gqli}. 

Photon production from hot dense matter will also be affected by the change of
the properties of vector and axial-vector mesons in medium. 
Here we are interested in the effects of `dropping masses' of vector and 
axial-vector mesons in medium on photon production. 
To see these effects, we consider 
thermal production rates of photons from hadronic matter 
at a given temperature but with reduced masses. Since we assume that
the pion mass does 
not change in medium, the in-medium mass of the vector meson is never less 
than $2m_\pi\approx 300$ MeV in the following calculations. 

First we consider two scattering channels
for photon production in hot hadronic matter.
These reactions have been shown to be dominant sources of
photons with $E_\gamma<3$ GeV in hot hadronic matter.
In Fig. 1 we show the thermal production rate of photons 
in $\pi+\rho\to\pi+\gamma$ reactions with four different values 
of the vector-meson mass at $T=160$ MeV. We use 
Weinberg's mass formula for the effective masses of axial-vector mesons in 
medium, $m^*_{a_1}=\sqrt{2}m_\rho^*$. 
We can see that the production rate increases as the effective 
masses of vector and axial-vector 
mesons are reduced. We have more than an order-of-magnitude increase 
at $E_\gamma\approx 0.5$ GeV with $m_\rho=300$ MeV. The enhancement in 
production rates can be understood as there are more thermal vector mesons 
with reduced mass and the cross section increases with in-medium masses. 
The shape of the spectra does not change much, except a shift of the 
peak to the low-energy region.   
Fig. 2 shows the results for $\pi+\pi\to\rho+\gamma$ reactions.
Even though the absolute magnitude is slightly less than in the previous case
and these spectra are convex from below, we can reach a similar 
conclusion for $E_\gamma \geq 0.5$ GeV. We observe enhanced photon production 
rates with reduced effective masses of mesons.  

On the other hand, we have opposite results for the decay channels.
In Fig. 3 we show the effect of using the in-medium mass of mesons for
the reaction $\rho\to\pi+\pi+\gamma$. The thermal production rates 
are reduced with medium effects.
This is simply due to the reduced phase space of decay processes 
as the mass of vector mesons decreases. 
Since the contribution from the decay channels is relatively small 
compared to that from scattering channels, we neglect them in the following. 
   
From these calculations we conclude that the total thermal production 
rates of photons increase with reduced vector and axial-vector meson masses.  
The production rate may be enhanced by 1 to 2 orders of magnitude. 
Thus, it is very interesting to ask whether we can see such an enhancement in 
the observed photon spectra in high-energy nucleus-nucleus 
collisions. To answer this question, we need to include the space-time 
evolution of the hot dense matter produced in nuclear collisions.

\section{Medium effect in expanding hot dense matter}

In the previous section we have shown that reduced in-medium masses 
of vector and 
axial-vector mesons lead to an enhancement in the thermal production rate
of photons by 1 to 2 orders of magnitude. Since experimental
data show no such enhanced photon production, it is very 
interesting to see in a model whether these medium effects can be 
expected to survive in the observable photon spectra. 
In order to get the inclusive photon spectra for comparison to
experimental data, we need to integrate the production rates over the 
space-time development of the expanding hot dense matter,
\begin{equation}
E_\gamma{dN\over d^3p}=\int d^4x \biggl[E_\gamma{dN\over d^4xd^3p}
\left(E_\gamma,T,\mu\right)\biggr] \,\,\, ,
\end{equation}
where $T,\>\mu$ indicate the temperature and density (chemical potential) 
dependence of the photon production rates. 
Thus, observable spectra also depend on the details of space-time evolution, 
on initial conditions, and on properties of the phase transition expected 
in hot dense matter.

In our calculation we assume a boost-invariant longitudinal expansion of 
the hot dense matter and use a fluid-dynamical description 
for the expanding system. The evolution of the system is governed by 
the conservation laws for energy and momentum,
\begin{equation}
\partial_\mu T^{\mu\nu}(x)=0   \,\,\, .
\end{equation}
In the absence of dissipation, the energy-momentum tensor $T^{\mu\nu}$
is given by 
\begin{equation}
T^{\mu\nu}=(\epsilon+P) u^\mu u^\nu - P g^{\mu\nu}  \,\,\, ,
\end{equation}
where $\epsilon$, $P$ and $u^\mu$ represent energy density, pressure, 
and local four-velocity of the medium, respectively. 
Further, we assume that density effects on photon production 
other than effects of the effective 
masses of vector and axial-vector mesons are 
negligible \cite{gqli_photon}.

Using simple one-dimensional hydrodynamic expansion with Bjorken's 
initial conditions \cite{bjorken}, 
the transverse momentum distribution of photons
(transverse momentum spectrum) at central rapidity is given by
\begin{equation}
{dN\over dp_t dy}(y=0)=\pi R^2_A\int \tau d\tau\int d\eta
\biggl[E_\gamma {dN\over d^4xd^3p}\left(E_\gamma,T\right)\biggr] \,\,\, ,
\end{equation}
where $\tau$ is the proper time, $R_A$ is the transverse size of the system,
and $\eta$ is the space-time rapidity.
The proper time $\tau$ can be found as a function of temperature from the 
conservation of entropy as  
\begin{equation}
\tau s(T) = const. \,\,\, ,
\end{equation}
where $s$ denotes the entropy density.

The dynamical properties of the expanding system --- initial temperature, 
phase-transition temperature and the lifetime of the 
mixed phase if there is a phase transition ---
depend on the details of how the vector-meson mass is changed in medium. 
To see the quantitative effect of the reduced in-medium mass 
on the photon production we consider two extreme cases: 
one in which the bulk of the change in the effective mass happens close 
to the critical temperature of a first-order phase transition (A),
and another with a more gradual change, where 
the effective mass starts to decreases significantly already at
around a temperature $T=100$ MeV (B). The simple parameterizations for these 
two effective-mass scenarios are displayed in Fig. 4.
In both cases we 
assume that the vector meson mass is not less than twice the pion 
mass, $m_\rho^{*} > 300$ MeV, even at very high temperatures. 
For axial-vector mesons we assume that 
the relation $m^{*}_{a_1}=\sqrt{2} m^{*}_\rho$ holds for the entire evolution. 

Transverse expansion is neglected in the present work.
The effect of transverse expansion on thermal photon production
was found to be important for the apparent temperature of the 
spectra\cite{th_wa80}. However, the goal of this investigation is to see
whether there is a difference between a calculation with in-medium masses
versus one with free masses.
To isolate this effect a simple model without transverse 
expansion is sufficient. Similarly, a more refined treatment
of freeze-out\cite{nlf97} is unnecessary as it would influence both
calculations the same way.  

\subsection{No phase transition}

First we determine the initial temperature of the system. This can be
obtained from entropy conservation. 
If pions carry most of the system's entropy at freeze-out, entropy 
conservation relates the initial entropy density to the total
observed pion rapidity density (including charged and neutral pions).
Using the fact that the entropy/particle for an ideal massless boson
gas is 3.6 one obtains 
\begin{equation}
\pi R^2_A \tau_0 s(T_0) =3.6 {dN_\pi\over dy}   \,\,\,  ,
\end{equation}
where $\tau_0$ is the initial proper time and 
$T_0$ is the initial temperature. 
In this paper we use $\tau_0 = 1 $ fm/c.
Since we assume no phase transition, the initial entropy
is given by the hadronic content of the system. Treating the hadronic
matter as a non-interacting gas of hadrons, the entropy is given by
\begin{equation}
s_H(T)={1\over T}\biggl[\epsilon_H(T)+P_H(T)\biggr]  \,\,\, ,
\end{equation}
with
\begin{eqnarray}
\epsilon_H &=& \sum_i{g_i\over (2\pi)^3}\int d^3p{\omega_i\over \exp
(\omega_i/T)-1}  \,\,\, ,\cr
P_H &=& \sum_i{g_i\over (2\pi)^3}\int d^3p{p^2\over 3\omega_i
[\exp(\omega_i/T)-1]}  \,\,\, ,
\label{pressure}
\end{eqnarray}
where the sum is over various boson species the system is composed of, 
$g_i$ is the degeneracy factor, and $\omega_i=\sqrt{p^2+m_i^2}$.

For the central collisions of S+Au at 200 GeV/nucleon at the CERN SPS, 
we get $T_0 \approx 300$ MeV with free masses for vector and axial-vector 
mesons. We have assumed that the hadronic phase consists 
of $\pi,\>\eta,\>\rho,\>\omega, \mbox{and} \>a_1$ mesons.
This value of the initial temperature seems to be rather high for 
a hadronic phase.  
On the other hand, the initial temperature of the same system reduces 
to $T_0\approx$ 250 MeV assuming that the effective masses of vector
mesons in medium reach the value shown in Fig. 4 at high temperature. 

The temperature evolution, as the system cools down, 
is obtained from the entropy conservation equation 
\begin{equation}
\tau s(T)=\tau_0 s(T_0)  \,\,\, .
\label{temp_evol_hadron}
\end{equation}
In Fig. 5 we show the time dependence of the temperature with 
reduced in-medium masses for vector and axial-vector mesons
according to the scenarios (A) and (B) displayed in Fig.~4. 
The results are compared to the temperature evolution for the same 
system but with free masses.  
Interestingly, details of mass changes do not influence 
the temperature evolution very much. The initial mass of the vector meson
has the only important effect on the evolution.  
At the very early stage of the evolution $\tau<5$ fm/c, the temperature 
changes in the same way for both mass profiles (A) and (B).
When the effective mass changes suddenly as in (A), there is an
early time interval without significant 
temperature change, resembling a first order phase transition 
in its effects. 
The system continues to cool down until it reaches the freeze-out 
temperature. In this paper we take the freeze-out temperature $T_f$=100 MeV. 
As mentioned earlier, we have a lower initial temperature for a
system with reduced masses compared to the value obtained with free masses. 
Furthermore, we can see that, in the entire evolution,
the temperature of a system with reduced vector and axial-vector 
meson masses remains below the temperature of a system without 
medium effects on the effective masses. 

These two medium effects, 
lower temperature at the beginning and in fact throughout the evolution,  
play significant roles in shaping the photon spectra from 
expanding hadronic matter. We show the results for the inclusive 
photon spectra obtained with reduced masses (two different profiles) 
in Fig. 6,  
and compare them to the results obtained with free masses. 
The in-medium effective mass induces a decrease in the 
production of inclusive photons, 
which is opposite to the effects of reduced masses on the 
thermal production rates of photons in the previous section. 
For photons with $p_t>1$ GeV/c, total production is reduced 
by an order of magnitude with reduced in-medium masses.
This indicates the importance of the 
reduced initial temperature and of the  lower temperature of the 
system during the entire evolution compared to 
that of a system with free masses. Even though the thermal production rates 
increase with `dropping effective masses', 
the increase is more than compensated by the reduced temperature of the system.
The details of the mass evolution have no significant effect on the 
photon production since most photons are produced at the very early stage 
of the evolution when there is no difference between the two mass profiles. 

\subsection{First-order phase transition}

In contrast to the previous subsection, here
we assume that the strongly-interacting matter produced
in high-energy central collisions is in the
quark-gluon plasma phase at the early stage of the evolution. 
Further, we assume that the system transforms into the hadronic phase
through a first-order phase transition as it expands and cools down. 
In this case the initial conditions are determined by the properties of 
quarks and gluons. 
Including massless up and down quarks and gluons, we have for 
the initial temperature
\begin{equation}
T_0 =\left(3.6 {dN_\pi\over dy}{1\over\tau_0\pi R^2_A}
                 {1\over (4/3)(37\pi^2/30)}\right)^{1/3}  \,\,\, .
\end{equation}
For the system produced in S+Au collisions at the CERN SPS, 
using the value of $\displaystyle{dN_\pi \over dy}$ from the WA80 experiment,
we get $T_0=203$ MeV.

The critical temperature can be determined by the pressure balance
requirement,
\begin{equation}
P_H(T_c)=P_Q(T_c) \,\,\, .
\end{equation}
The pressure in the quark-gluon plasma phase is given by 
\begin{equation}
P_Q(T) = 37\left({\pi^2\over 90}\right) T^4-B \,\,\, ,
\end{equation}
where $B$ is the bag constant and we use $B=206$ MeV \cite{wong}. In the
hadronic phase the pressure is given by Eq. (\ref{pressure}) with in-medium 
masses. Thus, the critical temperature depends on the effective masses of 
vector and axial-vector mesons. As we include the reduced effective masses 
of mesons, we get higher values for the critical temperature, $T_c=160$ MeV,
compared to $T_c=150$ MeV obtained with free masses. 

In Fig. 7 we show the temperature evolution assuming a first-order phase 
transition. We consider the medium effect on the vector mesons with the two 
different profiles (A) and (B), and compare the results to 
the evolution obtained with free masses. 
The evolution of the temperature in the plasma phase is given by 
\begin{equation}
T(\tau)=T_0(\tau_0/\tau)^{1/3} \,\,\, ,
\end{equation}
until the system reaches the phase transition temperature at time 
$\tau_1=(T_0/T_c)^3 \tau_0$.

In the mixed phase the temperature does not change,
but the entropy changes by converting quark-gluon plasma to hadronic gas. 
The mixed phase continues to time $\tau_2$ which is given by 
\begin{equation}
\tau_2=r \tau_1  \,\,\, ,
\end{equation}
where $r$ is given by the ratio of the entropy densities in the
two phases at the critical temperature, $r=s_q(T_c)/s_h(T_c)$ \cite{dilepton}. 
Thus, the lifetime of the mixed phase also strongly 
depends on the vector and axial-vector meson masses in medium. We find that 
the lifetime of the mixed phase is considerably reduced in the `dropping 
mass' scenario. With free masses the mixed phase continues for 
$\Delta\tau_{mix}=20$ fm/c.
However, this time is
reduced to $\Delta\tau_{mix}=5$ fm/c when we use in-medium masses.
This implies that there might be no first-order phase transition as 
the effective masses of vector mesons are reduced further in medium, 
as discussed in Ref. \cite{no_phase}. 
Since the entropy difference between the two phases decreases
as the masses are reduced in the hadronic phase, the
system transforms very rapidly from the plasma phase to the hadronic phase.
This is why we have a very short time for the mixed phase with 
`dropping masses'. 

In the hadronic phase the evolution is given by Eq.(\ref{temp_evol_hadron}) 
with $T_c$ and $\tau_2$ instead of $T_0$ and $\tau_0$, respectively.
As shown previously, 
the details of the temperature evolution do not depend strongly on 
how the effective masses of vector mesons change in medium. The hadronic 
phase soon cools down and the temperature of the system becomes lower than 
that for the system with free masses throughout most of the evolution. 
As a result, the system reaches the 
freeze-out temperature
earlier than with free masses. The entire lifetime of 
the hot dense matter becomes short when we include the medium effects on the 
vector and axial-vector mesons. 

Now we consider medium effects on the inclusive photon spectra from 
hot dense matter. 
In Fig. 8 we show the results with medium effects when a phase transition is
assumed, and compare them to those obtained with free masses. 
The enhancement shown in the thermal production rates is compensated 
to some extent 
by the reducing effects from the changes in dynamic properties of
the expanding system. Due to the short lifetime of the mixed phase there 
is a slight reduction in photon production from the plasma component of the 
mixed phase. However, 
the dominant contribution comes from the hadronic phase and the hadronic 
component of the mixed phase. These hadronic contributions show a significant 
increase due to the reduced masses, even though there are reducing effects 
of the low temperature of the system and of the short lifetime of the mixed 
phase. This is because most of the hadronic contribution now comes from 
a high temperature.
Also, we can see a rather apparent 
difference between the two scenarios (A) and (B), and compared to the results 
obtained with no phase transition. 
We have more photons with scenario (A) than with scenario (B). 
This is because of the difference in cooling rates of the system. 
The cooling rate is given by 
\begin{equation}
{dT\over d\tau}\sim\left[{1\over [s(T)]^2}{ds(T)\over dT}\right]^{-1}. 
\end{equation}
It turns out that this cooling rate  is $\approx 3$ times slower with mass 
profile (A) than with mass profile (B) at the early stage of the expansion 
when the temperature decreases from $160$ MeV to $140$ MeV. 

Finally, we compare our results to the upper limits
obtained by the WA80 experiment in Fig. 9. 
We find that the purely hadronic scenario is excluded by these upper limits
in our simple model even though we have an improved result with 
reduced masses. This is due to the rather high initial temperature 
of the system. However, we cannot rule out that a 
hadronic model using more degrees of freedom and a more realistic 
equation of state can be made consistent with the experimental information 
\cite{raju}. 
Our results are consistent with the WA80 upper limits
if a first-order phase transition is assumed in the model\cite{th_wa80}.
The medium effects due to the `dropping masses' of vector mesons do not 
change these results much, except an increase in production rates 
for high $p_t$ photons.  
We carried out a similar comparison with the preliminary WA98 
data\cite{rhic}, reinforcing that the data
are inconsistent with a pure hadronic calculation, while
the assumption of a first-order phase transition leads to
a reasonable description even with medium effects.
The main conclusion of this article is that
the data are still well-described in our simple model approach 
(with a first-order phase transition) without 
further modification when the medium effects of reduced masses
are included.   

\section{Conclusion}

We have studied photon production from hot dense matter as a probe of the 
properties of hadrons in medium. We are interested in the 
effect of the reduced masses of vector and axial-vector mesons.
Such reduction has been 
suggested as a precursor phenomenon to the chiral phase transition. 
This has been very interesting recently, since the 
enhancement of low-mass dileptons observed in the CERES and HELIOS 
experiments at the CERN SPS can be well described assuming 
`dropping masses' of vector mesons. 

First we have calculated thermal production rates of photons from 
hadronic matter at a given temperature but with reduced masses. 
It has been shown that the reduced mass
increases the production rates by an order of magnitude at $T=160$ MeV for
both reactions $\pi\rho\to\pi\gamma$ and $\pi\pi\to\rho\gamma$. However, 
the contribution from decay processes of vector mesons decreases with reduced 
effective masses simply because of the shrinkage of the phase space. 

In order to compare with experimental data we have included the
space-time evolution of the system as well as the changes of masses in medium. 
For the evolution of hot 
dense matter we used a one-dimensional hydrodynamic expansion and 
considered cases with and without a phase transition. 
This model is clearly too simple to describe the complicated 
system produced in nuclear collisions, 
but is sufficient to understand, rather qualitatively, medium effects on 
photon production. 
We find that not only the thermal production rates of photons, but also 
the initial conditions and phase-transition properties of the system 
depend on the changes of the effective masses of vector and axial-vector 
mesons.  

Assuming no phase transition, we have a relatively low value for the
initial temperature with reduced masses, $T_0=250$ MeV, due to the 
dependence of the initial entropy density of the system on 
the effective masses of vector and axial-vector mesons. 
Moreover, the temperature of the system with reduced masses 
is lower than that of the system 
with free masses throughout the evolution, which leads to an earlier 
freeze-out in the system with reduced masses.
The inclusive photon spectra 
from hot dense matter produced in nuclear collisions are 
subject to two medium effects: 
enhancement from reduced masses of vector and axial-vector mesons and 
suppression from the changes in initial conditions and temperatures of the 
system. The suppression effects overwhelm enhancement and lead to a reduction 
of the spectra by an order of magnitude for high energy photons.

When we assume that there is a first-order phase transition in the system, 
the initial temperature is decided by 
the properties of quarks and gluons and not affected by 
the medium effects on the vector meson mass. 
The phase transition temperature and the time spent 
in the mixed phase, however, depend on the reduced meson masses. We find that 
the reduced effective masses change the phase transition temperature by 
about 10 MeV. The lifetime of the mixed 
phase strongly depends on the effective masses of vector mesons. 
It turns out that 
$\Delta\tau_{mix}\approx 5$ fm/c with $m_\rho= 300$ MeV at $T=T_c$,
which is considerably shorter than $\Delta\tau_{mix}\approx 20$ fm/c 
with free masses. 
These changes reduce photon production and compensate the enhancement of
thermal production rates with reduced masses. 
The final results, however, show slightly enhanced photon production 
since most photons are produced at the early stage of the 
hadronic phase in which the temperature is higher and the vector meson mass is 
smaller than in the system with free masses.

Comparison with experiments shows that our results are consistent with 
the experimental information 
assuming a first order phase transition, even though we include the medium 
effects from `dropping masses' of vector mesons. 

In conclusion, thermal production rates of photons strongly depend on 
the changes in effective masses of vector and axial-vector mesons. 
However, these medium effects also change many important aspects 
of the dynamics 
of the expanding system. When we include these effects together, 
it is very hard to distinguish 
medium effects in the inclusive photon spectra observed at CERN SPS.

\acknowledgments
We thank Terry Awes for discussions of the experimental situation.
This work was supported in part by DOE Grant DE-FG02-86ER40251.

\vfill\eject\newpage

\section*{Appendix} 

The thermal emission rates of photons have been calculated from relativistic 
kinetic theory. For a reaction $1+2\to 3+\gamma$,  
\begin{eqnarray}
E_\gamma{dN\over d^4xd^3p}&=&{{\cal N}\over 2(2\pi)^3}
\int{d^3p_1\over (2\pi)^32E_1}
    {d^3p_2\over (2\pi)^32E_2}
    {d^3p_3\over (2\pi)^32E_3} (2\pi)^4\delta^{(4)}(p_1+p_2-p_3-p)\cr
&&\qquad\qquad\quad
\times\vert\bar{{\cal M}_i}\vert^2 f_1(E_1)f_2(E_2)[1+f_3(E_3)] \,\,\, .
\end{eqnarray}
Here $(E^j,p^j)$ is the four-momentum of the $j^{\mbox{th}}$ particle, 
$(E_\gamma,p)$ stands for the photon, 
and the $f$'s are Bose-Einstein or Fermi-Dirac distributions for each particle.
${\cal M}_i$ denotes the scattering amplitude for the given reaction,
which is averaged over initial states and summed over final states
to obtain $\vert\bar{{\cal M}_i}\vert^2$. The overall 
degeneracy factor $\cal N$ depends on the particular process. It is convenient
to introduce invariant variables $s=(p_1+p_2)^2$ and $t=(p_1-p_3)^2$. After 
integrating over angles the integration reduces to four dimensions,
\begin{eqnarray}
 E_\gamma{dN\over d^4xd^3p}&=&{{\cal N}\over 16(2\pi)^7E_\gamma}
\int^{\infty}_{s_0}ds\int^{t_{max}}_{t_{min}}dt\vert\bar{{\cal M}_i}\vert^2
\int_{[A]}dE_1\int_{[B]}dE_2\cr
&&\qquad\qquad\quad
\times f_1(E_1)f_2(E_2)[1+f_3(E_1+E_2-E_\gamma)]{1\over \sqrt{aE_2^2-2bE_2+c}} 
\,\,\, ,
\end{eqnarray}
which can be integrated numerically. The details of notation are given 
in Ref.\cite{song_photon}.

In the following we show numerical parameterizations for thermal production 
rates of photons obtained with axial-vector mesons included\cite{song_photon}. 
A parameterization is found by expressing the rate as a function 
of $\exp(-E_\gamma/T)$ and of dimensionless parameters like $x=T/m_\pi$ 
and $y=E_\gamma/m_\pi$. For the reaction $\pi\pi\to\rho\gamma$ we have
\begin{equation}
E_\gamma{dN_{\pi\pi\to\rho\gamma}\over d^4xd^3p}
=T^2\exp(-E_\gamma/T)F_{\pi\pi\to\rho\gamma}(x,y) \,\,\, ,
\end{equation}
where
\begin{eqnarray}
F_{\pi\pi\to\rho\gamma} &=& \exp\biggl[-12.055+4.387x+(0.3755+0.00826x)y\cr
                        & & \qquad +(-0.00777+0.000279x)y^2\cr
                        & & \qquad +(5.7869-1.0258x)/y
                                   +(-1.979+0.58x)/y^2\biggr] \,\,\, .
\end{eqnarray}

We obtain a similar parameterization for the reaction $\pi\rho\to\pi\gamma$ 
given by 
\begin{equation}
E_\gamma{dN_{\pi\rho\to\pi\gamma}\over d^4xd^3p}
=T^2\exp(-E_\gamma/T)F_{\pi\rho\to\pi\gamma}(x,y) \,\,\, ,
\end{equation}
where
\begin{eqnarray}
F_{\pi\rho\to\pi\gamma} &=& \exp\biggl[-2.447+0.796x+(0.0338+0.0528x)y\cr
                        & & \qquad +(-21.447+8.2179x)/y
                                   +(1.52436-0.38562x)/y^2\biggr] \,\,\, .
\end{eqnarray}

Finally we consider the decay channels. For the reaction $\omega\to\pi\gamma$
we have the same result as in ref. \cite{photon}, since there is no correction 
due to the $a_1$ meson. For $\rho$ decay into $\pi\pi\gamma$, 
the correction due to the $a_1$ meson is rather small compared to the 
reactions discussed earlier. 
The parameterization is given by    
\begin{equation}
E_\gamma{dN_{\rho\to\pi\pi\gamma}\over d^4xd^3p}
=T^2\exp(-E_\gamma/T)F_{\rho\to\pi\pi\gamma}(x,y) \,\,\, ,
\end{equation}
where
\begin{eqnarray}
F_{\rho\to\pi\pi\gamma} &=& \exp\biggl[-6.295+1.6459x+(-0.4015+0.089x)y\cr
                        & & \qquad +(-0.954+2.05777x)/y\biggr] \,\,\, .
\end{eqnarray}

\begin{figure}[p]
\caption{ 
Medium effects on the thermal emission rates of photons 
for the reaction $\pi\rho\to\pi\gamma$. We use $T=160$ MeV and four 
different values for the in-medium
vector meson mass, $m_\rho =300,\>450,\>650,\>770$ MeV from top.}   
\end{figure}

\begin{figure}[p]
\caption{ 
Medium effects on the thermal emission rates of photons 
for the reaction $\pi\pi\to\rho\gamma$. 
The meaning of the curves is the same as in Fig. 1.}   
\end{figure}

\begin{figure}[p]
\caption{ 
Medium effects on the thermal emission rates of photons 
for the reaction $\rho\to\pi\pi\gamma$.
The meaning of the curves is the same as in Fig. 1.}   
\end{figure}

\begin{figure}[p]
\caption{ 
Effective masses of vector mesons in medium. 
We consider two different mass profiles, (A) and (B) as explained in the 
text.}   
\end{figure} 

\begin{figure}[p]
\caption{ 
Temperature evolution of the system with no phase transition. 
The dotted curve is the result obtained with free mass. 
The dashed curve is the result with in-medium masses for vector mesons 
in scenario (A) of Fig. 4.
The solid curve is the result with in-medium masses for vector mesons 
in scenario (B) of Fig. 4.}   
\end{figure}

\begin{figure}[p]
\caption{             
Photon yield for a central $^{32}$S+Au collision at 200A GeV 
with no phase transition. The meaning of each curve is the same as in Fig. 5.} 
\end{figure}

\begin{figure}[p]
\caption{ 
Temperature evolution of the system with a first-order phase transition.
The meaning of each curve is the same as in Fig. 5.}   
\end{figure}

\begin{figure}[p]
\caption{
Photon yield for a central $^{32}$S+Au collisions 
at 200A GeV with a first-order phase transition. 
The meaning of each curve is the same as in Fig. 5.}   
\end{figure}

\begin{figure}[p]
\caption{ 
Comparison with the WA80 upper limits. 
Dashed and dash-dot curves are for the results obtained in pure 
hadronic phase (with no phase transition)  
without and with medium effects (A), respectively.
Dotted and solid curves are for the results obtained assuming a first-order 
phase transition without and with medium effects (A), respectively.}
\end{figure} 

\vfill\eject

\end{document}